\documentclass[a4paper,twoside]{article}
\usepackage[authoryear]{natbib}
\bibpunct{(}{)}{;}{a}{}{,}
\usepackage{graphicx}
\date{} 
\title{\large\bf\flushleft Chemical Abundances of the S star GZ Peg}
\author{\parbox{\textwidth}{\flushleft
\vspace{-0.5cm}
{\it L. Pomp\'eia}\\
\vspace{0.4cm}
{\small IP\&D, Universidade do Vale do Para\'iba, Av. Shishima Hifumi, 2911, S\~ao,
          Jos\'e dos Campos, 12244-000 SP, Brazil, Email: pompeia@univap.br}}}

\begin{document}
\twocolumn[
\begin{changemargin}{.8cm}{.5cm}
\begin{minipage}{.9\textwidth}
\vspace{-1cm}
\maketitle
%
%
\small{\bf Abstract:}

The chemical compositions of stars from the Asymptotic Giant Branch are still 
poorly known due to the low temperatures of their atmospheres and therefore the
presence of many molecular transitions hampering the analysis of atomic lines.
One way to overcome this difficulty is by the study of lines in regions free from
molecular contamination. We have chosen some of those regions to study the chemical
abundance of the S-type star GZ Peg. Stellar parameters are derived from spectroscopic 
analysis and a metallicity of -0.77 dex is found. Chemical abundances of 8 elements 
are reported and an enhancement of $s$-process elements is inferred, typical to that
of an S-type star. 


\medskip
\medskip
\end{minipage}
\end{changemargin}
]
\small

\section{Introduction}
The Asymptotic Giant Branch (AGB) is one of the most important phases of stellar evolution due to the rich nucleosynthesis that occurs during this stage, and therefore AGB stars are crucial objects for studies of Galactic chemical evolution. During this epoch stars produce 
important amounts of carbons and $s$-process elements which are brought to their atmospheres
by deep dredged-up events and then ejected to the interstellar medium by stellar winds. 
Although fundamental stones for the understanding of the final phases of low-mass stellar evolution 
and the stellar populations abundances, the chemical composition of such stars are still 
poorly known due to their low temperatures and to the blanketing of the atomic lines by molecular transitions. Some authors have tried to overcome this problem by selecting spectral regions free from molecular contaminations or the so called “molecular windows” (e.g. Smith \& Lambert 1985, 1990, Lambert et al. 1995, Vanture \& Wallerstein 2002a,b, 2003), and the use of the near-infrared and infrared regions (e.g. Origlia \& Rich 2005, Cunha \& Smith 2006, Woolf \& Wallerstein 2005a,b). Although working with a smaller number of lines, the chemical picture of these stars can still be delineated, provided that a careful inspection of the continuum is made and an appropriate model atmosphere is adopted. The knowledge of the elemental abundances of such stars is an important tool for the understanding of the many steps followed by red giants along the M-MS-S-C spectral sequence, and their role in the chemical evolution of galaxies.

In the present work we report the chemical abundances of GZ Peg, a S-type star. S stars show molecular transitions of ZrO and LaO, indicating an important enrichement of $s$-process elements. Although rich 
in those elements, lines of the unstable $s$-element $^{99}$Tc (which has a half-life 
of $\sim$ 10$^{6}$ years) are not observed in GZ Peg spectrum (Lebzelter \& Hron 1999). GZ Peg is therefore classified as an $extrinsic$ S-star, which is predicted to have inherited its $s$-process content in a previous mass-tranfer event from an AGB compagnion. In $intrinsic$ 
S giants, lines of Tc are observed and they are believed to have recently formed $s$-process elements and dredged-up to their atmospheres. The improvement of the quality of the stellar spectra and of the atomic constants and the availability of new model atmospheres allow now a robust analysis of the chemical patterns of such stars. This is the first of a series of papers about the detailed chemical abundance of $s$-process enriched stars. In this series the following questions are adressed:
(i) What is the detailed chemical pattern of the sequence M-MS-S stars? (ii) Are there S stars with different $s$-process contamination as seen in Ba-stars? And if yes, what are the factors determining the intensity of $s$-process contamination?; (iii) Does the stellar population to which the AGB star belongs affects the $s$-process production? If yes, what are the effects of this relationship?

\section{Observations and Analysis}
Observations were carried out at the 1.52m telescope of ESO 
(European Southern Observatory), La Silla, in September 1999. 
The spectra were obtained using the FEROS spectrograph (Fiber-fed Extended Range Optical 
Spectrograph) with wavelength range 356 to 920 nm and a resolution of R = 48,000. 
Reductions were performed using the DRS (online data reduction system 
of FEROS). A subsequent reduction was performed using the CONTINUUM, 
RVIDLINES and DOPCOR tasks of the IRAF package. The mean signal to noisy 
ratio of the spectrum is S/N = 61.97.

Reddening has been estimated according to Chen et al. (1998) and bolometric 
correction has been interpolated from the grids of Lejeune et al. (1997).
GZ Peg is a semirregular (SRA) variable according to the General Catalog of 
Variable stars (GCVS, Samus et al. 2004). The light curve of the star at AAVSO 
(American Association of Variable Star Observers) has been checked and a visual 
magnitude range of $\pm$ 0.4 is found for this star, therefore an uncertainty of
0.4 mag is expected. We have adopted the bolometric luminosity of M$_{bol}$ = -5.02
from Guandalini \& Busso (2008). The 
temperature was inferred by the IRFM using the empirical calibrations 
of Alonso et al. (1999) and 2MASS colors. The resulting values are T(V-K) = 
3502 K, T(J-H) = 3621 K and T(J-K) = 3454 K. These temperatures have been checked by the 
excitation equilibrium of the Fe~I lines (Fig. 1), and a best fit has been found for 
T$_{\rm eff}$ = 3600K. Although S stars are predicted to have masses in a range of 1.5 - 2 
M$_{\odot}$ (Jorissen et al. 1998), GZ Peg is an extrinsic S star, therefore a series
of tests have been made to better constrain the mass of this star. Tests requiring simultaneously 
(i) the excitation equilibrium of the Fe~I lines and (ii) the curve of growth fitting, were made with masses of 3 M$_{\odot}$ to 1.5 M$_{\odot}$ and the iteration procedure converged for a model with 2 M$_{\odot}$. This value has been adopted to derive the trigonometric surface gravity also using the Hipparcos paralaxe.

The stellar parameters have been inferred by requiring the ionization equilibrium of the Fe~I 
and Fe~II lines to derive the spectroscopic gravity and metallicity, and zero slope in the [Fe/H] vs. EW (equivalent width) plot to infer the microturbulence velocity. The resulting stellar parameters are the following: T$_{\rm eff}$ = 3600 K, log g = 0.06, [Fe/H] = -0.77 dex and $\xi$ = 1.88 kms$^{-1}$. The [Fe~I/H] vs. EW relation is depicted in Fig. 2 and the curve of growth fitting is given in Fig. 3. The model atmospheres adopted here is a new version of the MARCS models atmospheres (Gustafsson et al. 2008). We have selected 2 M$_{\odot}$ spherical models and interpolated between grids to obtain the best result.

A change of 50K in the temperature yields a change in gravity of 0.025 dex
and a change of 0.4 mag in the bolometric magnitude yields a change in gravity of
0.16 dex. We have determined the errors in stellar abundances due to stellar 
parameters uncertainties taking into account a change of $\Delta$ log g = +0.2 dex in 
gravity, $\Delta$ T$_{\rm eff}$ = +50 K in temperature, $\Delta$ [Fe/H] = +0.15 dex in 
metallicity and $\Delta$ $\xi$ = 0.10 kms$^{-1}$ in microturbulence velocity. The 
resulting errors are given in Table \ref{err}. The uncertainties have been added in 
quadrature and the total errors are given in the last column.

\begin{table*}[h*]
\begin{minipage}{140mm}
\begin{center}
\caption{Changes in chemical abundances due to stellar parameters uncertainties. In the last 
column we give the total error.}\label{err}
  \begin{tabular}{lccccc}
  \hline
Element & $\Delta$ log g   & $\Delta$ T$_{\rm eff}$ & $\Delta$ [Fe/H] & $\Delta$ $\xi$ & Total error \\
\hline
         & +0.2 dex & +50 K  & + 0.15   dex    & 0.10 kms$^{-1}$ & $\delta$ \\
\hline
  Fe 1    &  0.09   &  -0.03   &     -0.01     &  -0.05   &   0.11   \\
  Ti 1    &  0.05   &   0.04   &     -0.03     &   0.00   &   0.10   \\
  Ni 1    &  0.07   &  -0.05   &     -0.03     &   0.00   &   0.09   \\
  Cr 1    &  0.04   &   0.02   &     -0.01     &   0.00   &   0.07   \\
  Zr 1    &  0.07   &   0.07   &     -0.02     &   0.00   &   0.11   \\
  Nd 2    &  0.08   &  -0.02   &      0.08     &   0.00   &   0.11   \\
   Y 2    &  0.09   &  -0.02   &     -0.05     &   0.00   &   0.10   \\
  Ce 2    &  0.09   &   0.01   &      0.05     &   0.00   &   0.10   \\
  Ba 1    &  0.02   &   0.05   &      0.01     &   0.00   &   0.05   \\
\hline
\end{tabular}
\medskip\\
\end{center}
\end{minipage}
\end{table*}

\begin{table}[h]
\begin{center}
\caption{Chemical Abundances}\label{abund}
\begin{tabular}{lccc}
\hline
Element & [X/H]  & $\delta$$^a$ & Number of lines \\
 \hline
   Fe 1    & -0.77 & 0.04  &  13 \\
   Ti 1    & -0.42 & 0.11  &  9  \\
   Ni 1    & -0.62 & 0.13  &  6  \\
   Zr 1    & -0.19 & 0.11  &  11 \\
   Cr 1    & -0.28 & 0.32  &  2  \\
   Nd 2    &  0.35 & 0.13  &  5  \\
    Y 2    & -0.10 &  -    &  1  \\
   Ce 2    &  0.56 & 0.27  &  2  \\
   Ba 1    & -0.05 &  -    &  1  \\
\hline
\end{tabular}
\medskip\\
$^a$ Statistical error $\sigma$/$\sqrt{N}$ \\
\end{center}
\end{table}

\begin{figure}[h*]
\begin{center}
\includegraphics[scale=0.3, angle=270]{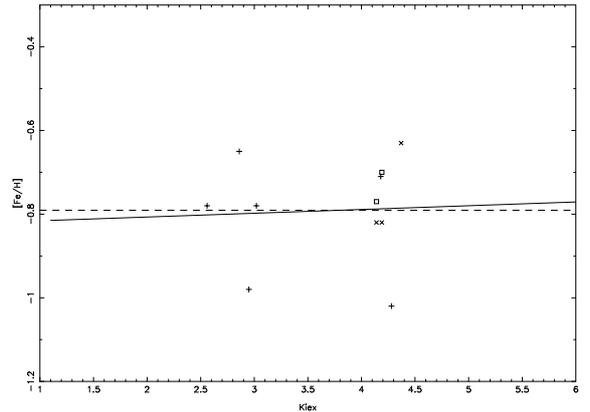}
\caption{Excitation equilibrium of the Fe~I lines for GZ Peg.}\label{exc}
\end{center}
\end{figure}

\begin{figure}[h*]
\begin{center}
\includegraphics[scale=0.3, angle=270]{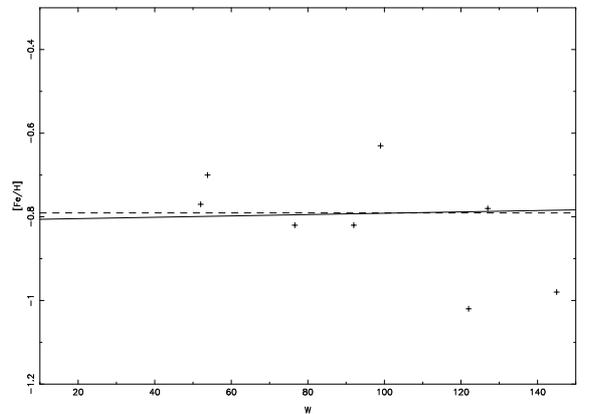}
\caption{[Fe/H] vs. EW of the Fe~I lines for GZ Peg.}\label{ew}
\end{center}
\end{figure}

\begin{figure}[h*]
\begin{center}
\includegraphics[scale=0.35, angle=270]{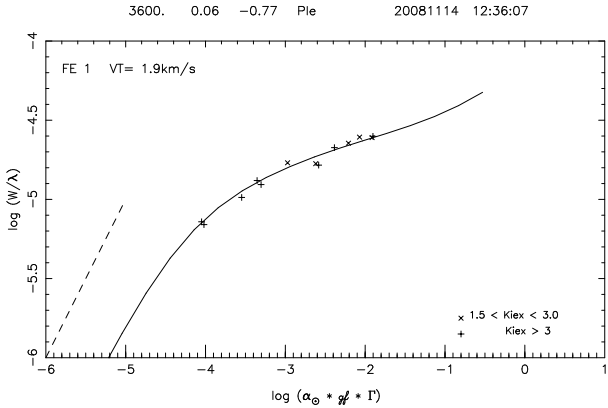}
\caption{Curve of growth fitting.}\label{curve}
\end{center}
\end{figure}

Chemical abundances have been inferred by the analysis of the equivalent widths
of the lines. The line list and the EW of the lines are given in Table \ref{line_list}.
In Table \ref{abund} we give the final abundances with the number of lines used in the 
analysis and the statistical errors ($\sigma$/$\sqrt{N}$). 

\section{Discussion}
\begin{figure*}[h]
\begin{center}
\includegraphics[scale=0.4, angle=0]{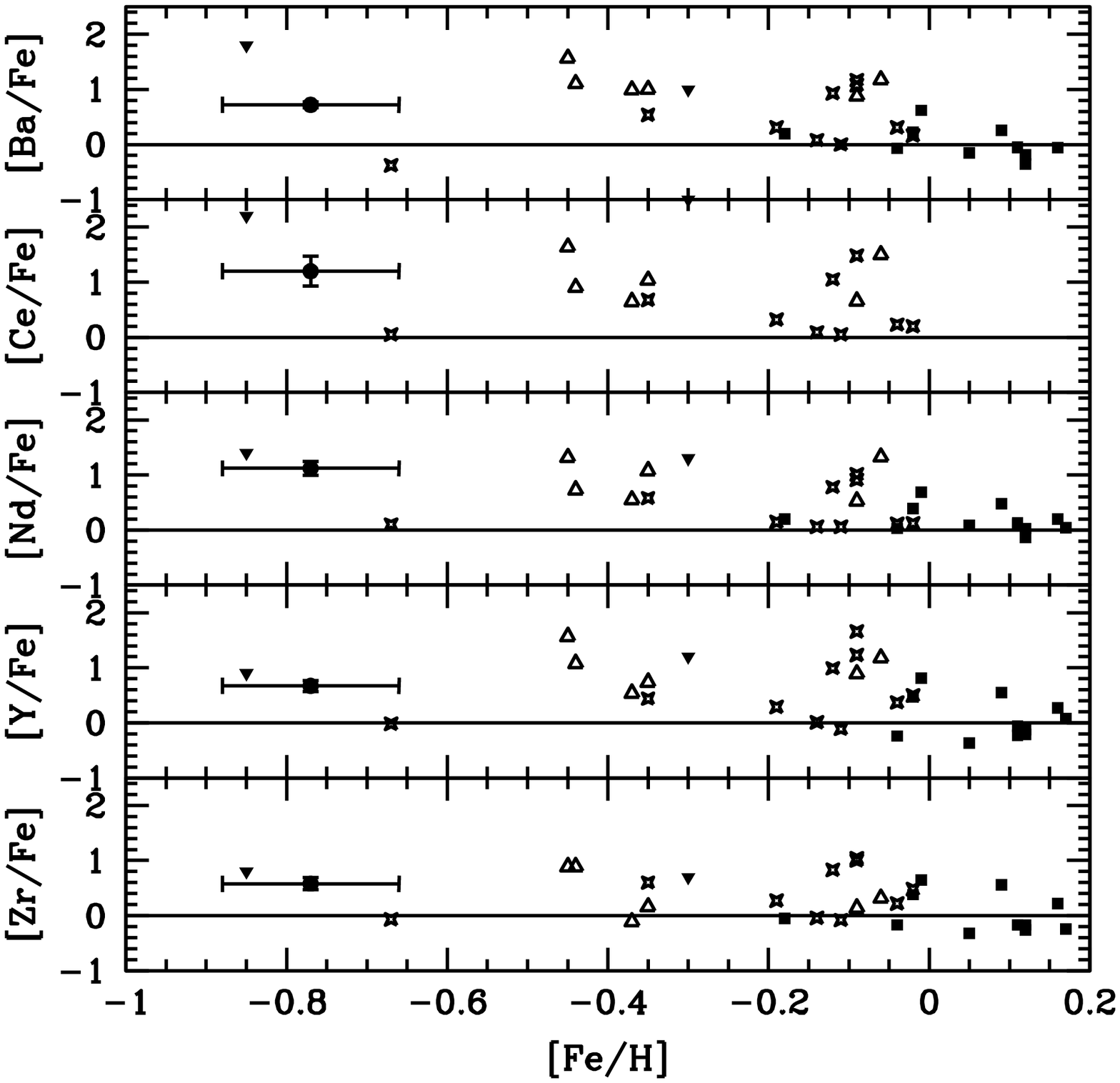}
\caption{Comparison of the chemical abundances of GZ Peg with MS, S and Ba
stars. The reference of the symbols are the following: filled circle - GZ Peg,
downward pointing triangles - S stars from Vanture \& Wallerstein (2003), 
open stars - MS and S stars from Smith \& Lambert (1986), squares - Ba-stars 
from Smiljanic et al. (20007). For the error bars we have adopted the highest 
error value between Tables \ref{abund} and \ref{err}. }\label{abundances}
\end{center}
\end{figure*}

\begin{figure}[h*]
\begin{center}
\includegraphics[scale=0.4, angle=0]{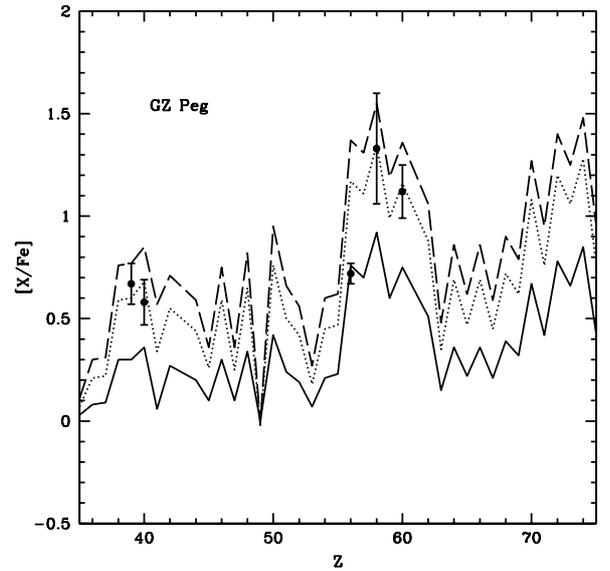}
\caption{[X/Fe] vs. Z for GZ Peg compared to theoretical surface abundances 
predicted for AGB stars of Goriely \& Mowlavi 2000. Our data are compared to 
models with [Fe/H] = -0.70 and M=2.5M$_{\odot}$. The solid line represents 
surface abundances predicted for 10 dredge-ups, the dotted line for 30 dredge-ups, 
and the dashed line for 50 dredge-ups. Error bars are the same as in Fig. 
\ref{abundances}.}\label{curve}
\end{center}
\end{figure}

A somewhat low metallicity of [Fe/H] = -0.77 has been found for GZ Peg. Abundances of [Ti/Fe] = +0.35 and [Ni/Fe] = +0.15 are derived, and a slightly high value for Cr, [Cr/Fe] = +0.48 is found (although this element is represented only by two lines with a high scater between them). For the $s$-process elements, we have Zr I and Nd II abundances better defined (with a larger number of lines) than Y II, Ba I and Ce II. We have found $s$-process abundances typical of a S star for GZ Peg. In Figure \ref{abundances} we plot our data with those of two S stars from Vanture \& Wallerstein (2003), MS and S stars from Smith \& Lambert (1986) and Ba-stars of Smiljanic et al. (2007) and Allen \& Barbuy (2006). As can be seen from this figure, GZ Peg shows supersolar abundance ratios for [Ba/Fe], [Zr/Fe], [Y/Fe], [Nd/Fe] and [Ce/Fe], and an enhanced pattern for [Nd/Fe] and [Ce/Fe] when compared to most of the s-enriched stars. The overabundance level is similar to that of the intrinsic S star S Uma (Vanture \& Wallerstein 2003), which has a metallicity of [Fe/H] = -0.3.

In order to better understand the origin of the abundance pattern of GZ Peg, we have compared 
in Fig. \ref{curve} the resulting abundances with those predicted by Goriely \& Mowlavi (2000) 
for an AGB star with metallicity [Fe/H] = -0.70 and mass of 2.5 M$_{\odot}$ which has undergone 10, 30 and 50 dredge-ups. 
Surface abundances predicted for 10 dredge-ups are depicted by the solid line, for 30 dredge-ups by the dotted line, and for 50 dredge-ups by the dashed line. Except for the Ba abundance, which has been derived from only one line, GZ Peg data are better described by the 30 dredge-up model. Such analysis may indicate that this star has probably accreted matter from an AGB star which has 
passed through an important number of dredge-up events. 

\section{Conclusions}

A detailed analysis of the extrinsic S star GZ Peg has been performed. The spectroscopic stellar
parameters derived for this star are: T$_{\rm eff}$ = 3600 K, log g = 0.06, [Fe/H] = -0.77
dex and $\xi$ = 1.88 kms$^{-1}$. The derived $s$-process abundances are in agreement with those of
S-type stars or other stars with enhanced $s$-process abundances. Comparing the abundance level
of GZ Peg with theoretical surface abundances of AGB stars from Goriely \& Mowlavi (2000)
we have found a good agreement with the 30 dredge-up model. Such analysis indicates
that the pristine AGB star which contaminated GZ Peg has passed through a considerable
number of dredge-up events.

\section*{Acknowledgments} 
This work has made use of the SIMBAD, VALD and AAVSO databases.

\begin{table*}[h*]{}
\begin{center}
\caption{Line List - Central wavelength of the lines, chemical element, excitation potential, oscilator strength, atomic data reference and equivalent width.}\label{line_list}
\begin{tabular}{lccccc}
\hline  
Wavelength & Element & $\chi_{exc}$ & log gf & Reference & EW \\
 \hline 
7411.180  &  FE1  &  4.283  &  -0.299   &  03  &   122.0  \\
7418.670  &  FE1  &  4.143  &  -1.226   &  03  &    92.0  \\
7443.002  &  FE1  &  4.186  &  -1.405   &  03  &    76.6  \\
7453.997  &  FE1  &  4.186  &  -1.907   &  03  &    53.8  \\
7461.530  &  FE1  &  2.559  &  -3.281   &  03  &   127.0  \\
7498.530  &  FE1  &  4.143  &  -1.940   &  03  &    52.0  \\
7511.040  &  FE1  &  4.178  &   0.229   &  03  &   188.0  \\
7531.170  &  FE1  &  4.371  &  -0.939   &  03  &    99.0  \\
7583.800  &  FE1  &  3.018  &  -1.990   &  03  &   161.0  \\
8611.800  &  FE1  &  2.850  &  -1.926   &  01  &   213.0  \\
8621.600  &  FE1  &  2.950  &  -2.321   &  01  &   145.0  \\
8674.750  &  FE1  &  2.830  &  -1.800   &  01  &   213.0  \\
8838.430  &  FE1  &  2.860  &  -2.050   &  01  &   200.0  \\
7337.780  &  TI1  &  2.239  &  -1.517   &  03  &   119.6  \\
7474.940  &  TI1  &  1.749  &  -2.187   &  03  &   102.6  \\
7496.120  &  TI1  &  2.236  &  -0.937   &  03  &   123.6  \\
8024.840  &  TI1  &  1.880  &  -1.140   &  01  &   159.0  \\
8334.390  &  TI1  &  0.820  &  -2.637   &  01  &   206.0  \\
8364.240  &  TI1  &  0.840  &  -1.756   &  01  &   319.2  \\
8377.860  &  TI1  &  0.830  &  -1.612   &  01  &   392.6  \\
8396.900  &  TI1  &  0.810  &  -1.779   &  01  &   380.0  \\
8412.360  &  TI1  &  0.820  &  -1.483   &  01  &   388.6  \\
8416.950  &  TI1  &  2.240  &  -1.034   &  01  &   133.7  \\
8417.470  &  TI1  &  2.120  &  -1.951   &  01  &   124.8  \\
8426.510  &  TI1  &  0.830  &  -1.253   &  01  &   468.0  \\
8438.920  &  TI1  &  2.260  &  -0.800   &  01  &   153.7  \\
8450.890  &  TI1  &  2.250  &  -0.903   &  01  &   106.8  \\
4385.240  &  NI1  &  2.740  &  -2.037   &  03  &    97.3  \\
7414.500  &  NI1  &  1.986  &  -2.549   &  03  &   182.7  \\
7422.300  &  NI1  &  3.635  &  -0.251   &  03  &   133.7  \\
7525.140  &  NI1  &  3.635  &  -0.546   &  03  &    91.6  \\
7555.600  &  NI1  &  3.847  &  -0.046   &  03  &   107.4  \\
\hline                                           
\end{tabular}                                    
\bigskip

Atomic References: \\                            
01 - Woolf \& Wallerstein (2005b) \\               
02 - Vanture \& Wallerstein (2002b) \\             
03 - VALD (http://ams.astro.univie.ac.at/~vald/) \\
04 - Vanture \& Wallerstein (2003) \\              
\end{center}                                     
\end{table*}

\setcounter{table}{2}

\begin{table*}[h*]{}
\begin{center}  
\caption{Continuation}
\begin{tabular}{lccccc}                          
\hline                                           
Wavelength & Element & $\chi_{exc}$ & log gf & Reference & EW \\
 \hline                                          
7574.043  &  NI1  &  3.833  & -0.533   &    03  &  104.3  \\
7437.138  &  CO1  &  1.956  & -2.876   &    03  &   64.6  \\
7553.963  &  CO1  &  3.952  & -0.955   &    03  &   43.1  \\
7355.935  &  CR1  &  2.890  & -0.290   &    01  &  208.0  \\
7400.226  &  CR1  &  2.900  & -0.110   &    01  &  202.2  \\
7462.364  &  CR1  &  2.910  & -0.040   &    01  &  213.6  \\
8348.283  &  CR1  &  2.710  & -1.830   &    01  &  112.0  \\
8976.880  &  CR1  &  3.090  & -1.110   &    01  &  156.2  \\
7439.890  &  ZR1  &  0.543  & -1.810   &    03  &   92.9  \\
7479.532  &  ZR1  &  1.834  & -1.380   &    03  &   24.8  \\
7551.493  &  ZR1  &  1.584  & -1.360   &    03  &   36.9  \\
7554.780  &  ZR1  &  0.520  & -2.280   &    03  &  111.6  \\
7562.129  &  ZR1  &  0.623  & -2.710   &    03  &   46.9  \\
8070.115  &  ZR1  &  0.730  & -0.790   &    01  &  141.1  \\
8133.011  &  ZR1  &  0.690  & -1.130   &    01  &  178.8  \\
8212.577  &  ZR1  &  0.650  & -1.320   &    01  &  167.2  \\
8305.987  &  ZR1  &  0.620  & -1.660   &    01  &  131.4  \\
8389.491  &  ZR1  &  0.600  & -1.760   &    01  &  148.7  \\
8584.281  &  ZR1  &  1.875  & -1.320   &    03  &   44.2  \\
7427.416  &  ND2  &  1.490  & -1.380   &    03  &   45.0  \\
7481.286  &  ND2  &  0.205  & -2.900   &    03  &   30.8  \\
7577.496  &  ND2  &  0.205  & -2.930   &    03  &   54.0  \\
8594.883  &  ND2  &  1.140  & -1.860   &    03  &   33.9  \\
8643.478  &  ND2  &  1.200  & -1.750   &    03  &   42.3  \\
7947.597  &  RB1  &  0.000  & -0.170   &    01  &  104.6  \\
7450.276  &   Y2  &  1.748  & -1.590   &    03  &   43.8  \\
8810.854  &  CE1  &  0.302  & -1.665   &    03  &   18.4  \\
7486.557  &  CE2  &  2.548  & -0.394   &    03  &   23.5  \\
8772.135  &  CE2  &  0.357  & -2.515   &    03  &   75.4  \\
7488.077  &  BA1  &  1.190  & -0.230   &    04  &   29.3  \\
\hline                        
\end{tabular}                 
\bigskip                      
\end{center}                  
\end{table*}                  
                              
\end{document}